# Second order front tracking algorithm for Stefan problem on a regular grid

Robert D. Groot[*]
Unilever Research Vlaardingen,
*P.O. Box 114, 3130 AC Vlaardingen, The Netherlands*

A brief review of the Stefan problem of solidification from a mixture, and its main numerical solution methods is given. Simulation of this problem in 2D or 3D is most practically done on a regular grid, where a sharp solid-liquid interface moves relative to the grid. For this problem, a new simulation method is developed that manifestly conserves mass, and that simulates the motion of the interface to *second order* in the grid size. When applied to an isothermal simulation of solidification from solution in 1D at 50% supersaturation for only 5 grid points, the motion of the interface is accurate to 5.5%; and for 10 points the result is accurate to 1.5%. The method should be applicable to 2D or 3D with relative ease. This opens the door to large scale simulations with modest computer power.

## 1   Introduction

Melting and freezing from solution or from a pure liquid are examples of moving boundary problems. These problems are described by time-dependent differential equations that need to satisfy boundary conditions at an interface, where the position of this interface is not known at forehand but has to be determined as part of the solution [1]. This class of problems is known as Stefan problems, after Josef Stefan (1835-1893) who studied the melting of polar ice in 1889 [2].

Although the problem is over a century old it still draws a lot of attention, firstly because of the large number of practical applications, and secondly because solutions are difficult to obtain. Applications are in the fields of biology, where e.g. ice formation can induce cryo-injury in freezing cells [3, 4]. Other applications are in food systems, in particular in the dendritic growth of ice in sugar solutions [5, 6], and further applications are in metal welding [7, 8, 9]. An authoritative review of the field has been given by Langer [10], a more recent review is by Asta *et al.* [11].

When crystals grow in conditions far from equilibrium, growth can become unstable, giving rise to pattern formation [10]. Materials for which the solid-fluid interfaces are rough at the molecular level but smoothly rounded on a macroscopic scale, growth is rapid. For such substances, growth is controlled by the diffusion fields in the neighbourhood of the solidification front only. Substances in this category include most metals and alloys and some organic crystals. Ice is an intermediate case, in which slowly growing facets occur parallel to the basal plane, but surfaces are rounded and grow rapidly in the hexagonal directions [10].

When such a solid grows into a supercooled melt or solution, a flat interface is generally unstable. Undulations form at one particular wavelength, leading to dendritic or cellular growth. The early stages of such instability have been observed for ice growing from a salt solution by Körber and Rau [3]. Dendritic ice growth for pure water was studied by Shibkov *et al.* [12], and from sugar solutions by Butler [5,6].

The onset of such growth instabilities was first studied theoretically by Mullins and Sekerka for a flat interface [13], and for a spherical crystal growing into a supersaturated solution [14]. Trivedi and Kurz [15] reanalysed the stability problem and extended the growth instability criterion to large thermal Péclet numbers, or growth velocities. A major step forward in the understanding of capillarity effects and the selection of the tip radius and velocity has come from Langer and Müller-Krumbhaar [16, 17], who predicted a universal growth law for dendritic growth rates, which was tested experimentally [18].

These results demonstrate detailed understanding of the problem of crystal growth instability and resulting morphology for certain cases of idealised geometry. However, when it comes to solid morphology in cases where several dendrites grow simultaneously, or form a three-dimensional network, analytic theory cannot be used. To analyse those cases, we need to resort to numerical methods. Several methods are used in the literature. Broadly speaking, these are the phase-field method [19], front tracking methods [20] and the level set method [21]. Each method has their merits and disadvantages.

The phase-field method is defined on a regular grid. The solid-liquid interface is smeared out over several grid points, which eliminates singularities in the equation of motion, but it requires many grid points and hence large computing power. The level set and front tracking methods simulate a sharp interface, which eliminates unphysical artefacts associated with a wide interface. Moreover, boundary conditions at (moving) boundaries can be implemented naturally. The down side is that without precautions, volume or mass are not conserved [22], which prompts the use of very small grid cells, or adaptive grid refinement. This makes the level set method either more computationally intensive than phase-field, or more complicated.

---

[*] Corresponding author.
  *E-mail address*: rob.groot@unilever.com (R.D. Groot).

Both in level set and in front tracking methods boundary conditions for the Stefan problem are enforced in a physical way, by determining the field gradient at the interface. The resulting finite difference scheme may be globally of second order in grid spacing and time step [20], but in current implementations the update equation for the phase front is of first order in the grid spacing. This limits the application to rather shallow concentration or temperature fields near the interface, which in turn forces the use of relatively fine grids. Here we set out to track the root cause why front updating is only of first order in the grid size, and how this is linked to volume non-conservation. To this end we concentrate on a 1D system only, but it is anticipated that the resulting scheme can be used in higher dimensions as well. With a second order scheme to update the phase front a coarser grid can be used with the same accuracy as in present first order schemes. This in turn opens the way to large scale simulations with modest computing power.

This article is organized as follows. In the next section, we give a brief summary of the Stefan problem, and a short review of the solutions methods is given in Section 3. Section 4 presents an error analysis of various methods to determine the first and second derivatives of a concentration or temperature field at or near an interface, and the mass conservation problem is analysed in Section 5. Based on this, a second order mass conserving update scheme is developed in Section 6, and conclusions are summarized in Section 7.

## 2 The Stefan problem

The generic problem is depicted in Fig. 1. To make the presentation more concrete, we will refer to the solid as ice and to the liquid phase as a concentrated solution, or matrix, although the same relations hold for other materials, like metals. Generally, temperature gradients in the solid phase are different from those in the liquid. Without solute (concentration $C$) the surface temperature $T_s$ equals the melting temperature of pure water, $T_s = T_0$. In that case the interface temperature is fixed, and the growth or melting rate of the ice phase follows from a heat balance.

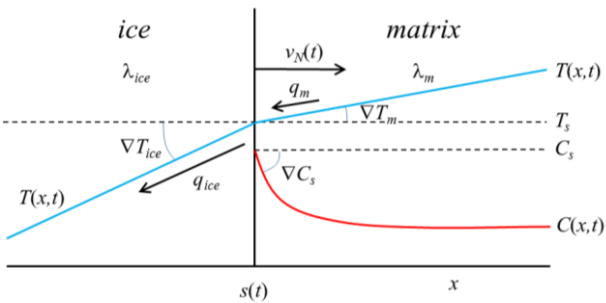

Fig. 1. Temperature distribution, heat flow and solute distribution near an ice-matrix interface.

The heat transport per unit of area in the matrix phase is $q_m = -\lambda_m \nabla T_m$, where $\lambda_m$ is the thermal conductivity of the matrix phase. Similarly, heat transport in the ice phase is given by $q_{ice} = -\lambda_{ice} \nabla T_{ice}$. Thus, the net amount of heat *extracted from* the interface per unit of area and time is given by $q_m - q_{ice} = \lambda_{ice}\nabla T_{ice} - \lambda_m \nabla T_m$. Freezing water produces latent heat $L$ per unit of mass. For normal growth rate $v_N$, the heat produced per unit of time and per unit of area is $L\rho_{ice}v_N$, where $\rho_{ice}$ is the density if ice. Produced heat must balance the extracted heat, hence the growth rate must satisfy

$$L\rho_{ice}v_N = \lambda_{ice}\nabla T_{ice} - \lambda_m \nabla T_m \qquad (1)$$

This is generally known as the (first) Stefan condition. Often the approximation is made that the density of the solid phase equals that of the pure solvent. When this approximation is made, all transport is diffusive.

In a binary solution, there is another boundary condition to be satisfied. Generally, the solute does not, or to a very limited amount, partition into the solid phase. Therefore, the solid can only move when the solute diffuses away. Analogously to the thermal diffusion as in Eq. (1), this condition is

$$(1-k)C_s v_N = D_{ice}\nabla C_{ice} - D_m \nabla C_s \approx -D_m \nabla C_s \qquad (2)$$

Here, $k$ is the solute partition coefficient, $C_s$ is the solute concentration at the liquid side of the surface, $C_{ice}$ is the solute concentration at the solid side of the surface, and $D_{ice}$ and $D_m$ are the Fick diffusion coefficients of the solute in the ice phase and in the matrix phase respectively. For most practical applications, when the solid is water ice, we can put $k = C_{ice} = D_{ice} = 0$.

A third boundary condition specifies the temperature (or solute concentration) at a curved solid-liquid interface. Due to the solid-liquid surface energy, the Laplace pressure leads to a shift in the melting point at the interface, which is known as the Gibbs-Thomson boundary condition. This is

$$T_s = T_m(C_s) - 2\Gamma/R - \beta v_N \qquad (3)$$

where $T_m(C_s)$ is the melting temperature for a *flat* interface at the *local* solute concentration $C_s$; $\Gamma$ is the Gibbs-Thomson parameter; and $R$ is the geometric mean radius of curvature (positive for curvature towards the solid, negative towards the liquid). For pure water $\Gamma = \gamma_{wi}T_0/\rho_{ice}L \approx 26 \cdot 10^{-9}$ K m, where $\gamma_{wi} = 0.029$ J m$^{-2}$ is the water-ice surface energy, $T_0 = 273.15$ K, $\rho_{ice} = 917$ kg m$^{-3}$ and $L = 333$ kJ kg$^{-1}$. See e.g. Van Westen and Groot for detailed thermodynamic parameters of the water-sucrose system [23]. Parameter $\beta$ is the kinetic coefficient, which describes a velocity-dependency of the interface temperature [19]. In most physical cases of diffusion-limited growth $\beta$ vanishes. The three equations Eq. (1–3) fix the surface velocity, the surface temperature and the surface concentration for a given set of temperature and concentration gradients.

Further to the boundary conditions, the Stefan problem is determined by the transport of heat and mass in the liquid and solid phases. When the solid density equals the density of the pure solvent, all transport is diffusive, and the transport equations are

$$\rho c_p \frac{\partial T}{\partial t} = \nabla \cdot (\lambda \nabla T) \qquad (4)$$

$$\frac{\partial C}{\partial t} = \nabla \cdot (D \nabla C) \qquad (5)$$

These equations apply to both the solid and the liquid phases. When there is a density difference between solid and pure solvent, solidification will induce flow in the liquid phase. Due to this flow, heat and mass transport is not only diffusive, but are also convective. To describe the effect of convection, one should replace the partial time derivative in the liquid phase by the convective derivative [24, 25]

$$\frac{D}{\partial t} = \frac{\partial}{\partial t} + \mathbf{u} \cdot \nabla \qquad (6)$$

where $\mathbf{u}$ is the velocity field in the liquid. This should be obtained from the Navier-Stokes equation, subject to appropriate boundary conditions at the solid-fluid interface. For an incompressible, Newtonian fluid these are [24, 25, 26, 27]

$$\begin{cases} \nabla \cdot \mathbf{u} = 0 \\ \rho \left( \frac{\partial \mathbf{u}}{\partial t} + \mathbf{u} \cdot \nabla \mathbf{u} \right) = -\nabla P + \mu \nabla^2 \mathbf{u} \end{cases} \qquad (7)$$

where $\rho$ is the liquid density, $P$ is the (isotropic) pressure and $\mu$ is the dynamic viscosity. Here, we concentrate on the 1D case, hence hydrodynamics will not be included.

An important parameter in the Stefan problem is the Lewis number, which is defined by the ratio of thermal diffusivity and solute diffusion:

$$\mathrm{Le} = \frac{\lambda}{\rho c_p D} = \frac{\alpha}{D} \qquad (8)$$

If $\mathrm{Le} \gg 1$ thermal diffusion is much faster than solute diffusion. In that case, the temperature profiles will be much wider than the concentration profiles. For freezing aqueous solutions, this is often the case. In the limit $\mathrm{Le} \to \infty$, the system is isothermal, and Eq. (1) and (4) become irrelevant. In the limit $\mathrm{Le} \to 0$, Eq. (2) and (5) become irrelevant, and ice growth is determined by thermal diffusion only. In practice, thermal diffusivity and solute diffusivity depend on both composition and temperature. Therefore, the transport equations are coupled and non-linear.

## 3 Solution methods

Very few analytical solutions are available in closed form. These are mainly one-dimensional cases of an infinite or semi-infinite region with simple initial and boundary conditions, and constant thermal properties. These solutions take on the form of functions of a single variable $x/\sqrt{t}$ (or $r/\sqrt{t}$ in spherical symmetry) and are known as similarity solutions [1]. These solutions are particularly useful to check the validity of numerical solution methods. An example for the isothermal case in a half space is given in the Appendix.

### 3.1 Moving grid method

In numerical work one often considers either isothermal (solute diffusion-limited) growth, or growth from a pure melt. In either case only one transport equation needs to be considered, but in general we need to solve both the temperature and the solute profiles. One method to solve these numerically is to use a moving grid [1]. In this method, the grid is deformed to match the position of the (sharp) solid-fluid interface. Since after each time step the interface always coincides with a grid point, the gradients at the interface can be determined straight-forwardly from numerical differentiation on a regular grid. In the isothermal case the concentration at the interface is fixed at its thermodynamic equilibrium value, hence the interface position at the next time step follows from the boundary condition Eq. (2) as

$$s(t + \delta t) \approx s(t) + \delta t D_m \frac{(3C_{eq} - 4C_1 + C_2)}{2C_{eq} h} \qquad (9)$$

Here, $s(t)$ is the interface position, $C_{eq}$ is the equilibrium solute concentration, and $C_1$ and $C_2$ are the concentrations at distances $h$ and $2h$ from the interface. In this approximation, the concentration gradient at the interface is obtained from a parabola fit through the known concentration values at the interface and the next two points.

After each time step the grid is re-adjusted to maintain a uniform grid starting at $s(t+\delta t)$. As the grid is adjusted, the concentration field at the new grid positions is interpolated from the old concentration field at the old positions. While the concentration at each point decreases in time, the interface has to move to maintain the mass balance. Unfortunately, a systematic error of order $O(h^2)$ is made in determining the concentration gradient at the interface. Errors thus accumulate and disturb the mass balance, unless precautions are taken.

A moving grid method was also used by Wollhöver *et al.* [4] to calculate the coupled temperature and concentration profiles in freezing salt water. They guess a position $s(t+\delta t)$ at the next time step and solve the concentration profile from the diffusion equation using a decomposition method by Vichnevetsky [28]. The concentration at the interface defines the local temperature, which is then used as a boundary condition to solve the temperature profile. In general, this solution will not satisfy the first condition Stefan condition Eq. (1). Hence a new positions are tried until both boundary conditions Eq. (1) and (2) are satisfied.

### 3.2 Implicit and explicit update schemes

A moving grid method is practical in one dimensional systems, but difficult to apply in higher dimensions. In 2D or 3D, the grid would need to deform with the solid, which makes it more difficult to solve the diffusion equation efficiently. There are two ways in general to solve the diffusion equation numerically, explicitly and implicitly. In an explicit solver, the profile at the new time step is written explicitly as function of the previous solution, e.g. if temperature $T_n$ is defined at grid positions $x_n = nh$, the temperature in the next step is obtained as

$$T_n(t+\delta t) = T_n(t) + \alpha \delta t [T_{n+1}(t) - 2T_n(t) + T_{n-1}(t)]/h^2 \quad (10)$$

This solves the heat equation explicitly, but unfortunately the solution is stable only for time steps

$$\delta t < \xi h^2 \quad (11)$$

where $\xi$ is a constant that depends on the problem. Things get worse when we solve a diffusion equation in combination with moving boundary conditions with curvature-driven growth [20]. It has been shown by Hou et al. [29] that the presence of capillarity terms in the interface evolution equation can lead to a severe numerical stability constraint if the interface is updated explicitly. In that case, the above criterion takes the form

$$\delta t < \xi h^3 \quad (12)$$

In Vichnevetsky's method mentioned above, the diffusion equation at the next time step is solved implicitly from the solution at the current time step. This is similar to the Crank-Nicolson scheme, which predates it by some 20 years [30]. Strictly speaking, both methods apply to linear differential equations only. The point of these update algorithms is that they are stable because they combine the (unstable) forward solution and the (stable) backward solution. Consequently, one can take big time steps without numerical instabilities. For the moving boundary problem this is a large advantage.

The Crank-Nicolson scheme can be applied in 1D, 2D and 3D systems, and involves the solution of a large ($N \times N$) matrix equation. The 1D update scheme is based on

$$\frac{T_x(t+\delta t) - T_x(t)}{\delta t} = \tfrac{1}{2}\alpha[\nabla^2 T_x(t) + \nabla^2 T_x(t+\delta t)] \quad (13)$$

Hence the mean time derivative between times $t$ and $t+\delta t$ is related to the average square gradient at the old and new time steps. Thus, the temperature at the next time step is defined implicitly and should be solved from

$$T_x(t+\delta t) - \tfrac{1}{2}\alpha\delta t \nabla^2 T_x(t+\delta t) = T_x(t) + \tfrac{1}{2}\alpha\delta t \nabla^2 T_x(t) \quad (14)$$

At this point the new temperature is expressed in the previous temperature field, but we still need to solve an ordinary differential equation. When the temperature field is represented on a regular grid, we obtain a matrix equation for an $N$-dimensional vector:

$$\sum_j A_{ij} T_j(t+\delta t) = F_i(t) \quad (15)$$

where the diagonal elements of matrix **A** are $A_{ii} = 1+\alpha\delta t/h^2$; the elements above and below the diagonal are $A_{i,i\pm 1} = -\alpha\delta t/2h^2$; and where **F**($t$) is a known vector following from Eq. (10) and (13). For insulated boundary conditions the first and last diagonal elements are $A_{11} = A_{NN} = 1+\alpha\delta t/2h^2$. This equation can be solved by straightforward matrix inversion. For the particular case that we have here, where all matrix elements are zero apart from those in a small band around the main diagonal, the amount of work to solve this equation is of order $O(N)$. In the example given in Eq. (15), where the matrix is *tridiagonal*, this can be done efficiently with the Thomas algorithm. With a tridiagonal matrix we can account for nearest neighbour interactions on the grid. To obtain more accurate results near boundaries we may use a pentadiagonal solver [31], so that we can include second neighbour interactions.

To solve the set of equations in Eq. (15) it is important that the matrix has limited band width. This can be guaranteed only if the temperature field is defined on a regular lattice. If in a 2D or 3D simulation the grid points are irregular, as in a co-moving grid, matrix inversion becomes inefficient. For this reason, it is highly desirable to solve the diffusion and heat equations on a regular Cartesian grid. To enable generalisation to 2D and 3D simulations we use a fixed grid, and allow the interface to move across the grid.

### 3.3 Fixed grid methods

#### 3.3.1 Phase-field method

There are two main ways to solve moving boundary problems on a fixed grid: the phase-field method and sharp interface methods. The main ingredient of the phase-field method consists in distinguishing between phases with a non-conserved order parameter, or phase field $\phi$, which is constant within each phase. This field varies smoothly across a spatially diffuse interfacial region of finite width $W$. Its dynamics is then coupled to that of the temperature field in such a way that the equations for the two fields reduce to Eq. (1), (3) and (4) in the so-called sharp interface limit of the model, introduced by Karma and Rappel [19, 32], where the interface is curved on a length scale much larger than $W$. Parameters must be tuned in a particular way by analysing the phase-field asymptotics, so that the Gibbs-Thomson condition, Eq. (3), is correctly satisfied. Otherwise an unphysical velocity-dependent term arises in the interface temperature, i.e. $\beta \ne 0$ in Eq. (3), whereas it should vanish for diffusion-limited growth.

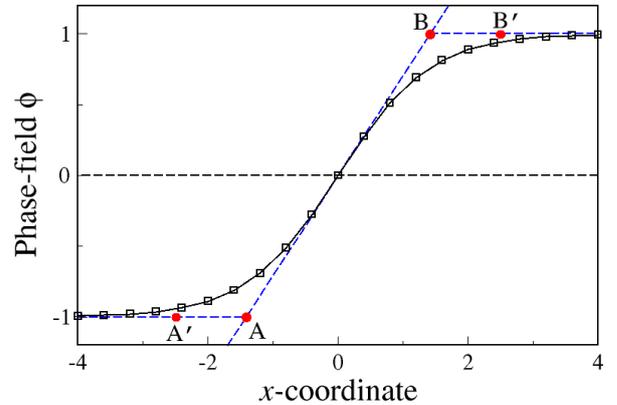

Fig. 2. Solid-liquid interface in phase field model for width $W = 1$ and grid size $h = 0.4$.

A typical example of a phase-field is [19, 32] $\phi = \tanh(x/W\sqrt{2})$, for $W = 1$ and grid size $h = 0.4$, as shown in Fig. 2. The square dots in Fig. 2 denote the grid points in the phase-field model. Although the width parameter is $W = 1$, the interface is actually smeared out over a wide area. For instance, the distance between points A and B in Fig. 2 is $2\sqrt{2} \approx 2.8$, and covers about 7 grid points. If we check from the graph where the phase-field runs horizontally, we may use points A' and B' in Fig. 2. The distance A'B' covers about 12 grid points. Thus, even though the sharp interface limit is a major step forward compared to older formulations, it still requires many grid points to describe just one interface. The condition of small curvature ($W \ll R$) aggravates the problem. In two and three dimensions this requires vast computational systems.

Another problem with the phase-field method arises for solidification from solution, when we deal with large Lewis numbers. In that case, the solute diffusion layer is very thin while the temperature profile is extended. Udaykumar and Mao [33] report that in their calculations of ice growing from salt solutions the solute boundary layers are extremely thin as compared to the width of the temperature field. Thus, unless very fine meshes are used, the width of the diffuse interface which is spread over a few mesh cells, can be comparable to the solute boundary layer thickness. This is clearly unphysical. Therefore, a sharp treatment of the interface is highly desirable in calculating the solidification of impure materials.

Finally, phase-field asymptotics for *unequal* diffusivities in the solid and liquid phases can be problematic [34]. Correction terms that are inconsistent with the sharp-interface equations are generated, and non-monotonic behaviour is required in the interfacial region. This requires extra grid resolution and hence slower computational performance. Generalization of the phase-field approach to handle discontinuous material properties requires a better understanding of the mapping between the phase-field model and the sharp-interface formulation to avoid problems.

*3.3.2 Front tracking methods*

Sharp interfaces can be simulated in various ways. One obvious way is to track the position of the interface explicitly. Examples are the immersed boundary method [20, 35, 36] and the immersed interface method [37]. In the immersed boundary method, introduced by Juric and Tryggvason [35, 38] the effect of the interface is transmitted to the field equation solver using smoothed delta functions and Heaviside functions, which show up as source terms in the transport equation. In this sense, the effect of the interface is still spread out over some grid cells. In the immersed interface method, the interface is mathematically sharp, and its effect is accounted for by including the physical boundary conditions.

Tracking the interface position is natural in a 1D system. This method was used by Wollhöver et al. [4] to calculate the coupled temperature and concentration profiles in freezing aqueous salt solutions. They used a moving grid to solve the transport equations. Udaykumar and Mao [33] reproduced their results with a fixed grid, using a second order discretization for the Poisson equation on irregular domains to impose the Gibbs-Thomson boundary condition [39]. This latter method can also be used in 2D and 3D systems [38].

In two dimensions the ice front is a curve, represented by interfacial points. These points are distributed at a distance of roughly the grid size. At each interfacial point, curvature follows from the relative positions of the neighbouring points. Together with the Gibbs-Thomson boundary condition (Eq. (3), with $\beta = 0$) this determines the temperature depression at each interfacial point. To find the interfacial temperature gradient, a normal to the interface is constructed. Normal probe points $n1$ and $n2$ are chosen at one and two grid distances away from the interface. The temperature at these points is determined by *bilinear* interpolation. Finally, the temperature gradient at the interface follows from a parabola fit through the temperatures at the interface marker and nodal probes $n1$ and $n2$ [20]:

$$\left(\frac{\partial T}{\partial n}\right)_{int} \approx -\frac{3T_{int} - 4T_{n1} + T_{n2}}{2h} \qquad (16)$$

Note the similarity with Eq. (9), where the same approach was used in 1D. The same method is used to determine the solute concentration gradients.

In a finite difference scheme to solve the diffusion equation, we need to obtain the square gradient of the concentration and temperature at each grid point. When a solid-liquid boundary cuts across the grid, a complication arises for the points that are close to the boundary. In those cases the field values at the interface are used to estimate the second derivatives in the adjacent field point.

From this point onwards, the method is similar to the method by Wollhöver et al. [4]. The normal velocity is solved from one boundary condition, Eq. (1), and the discretised temperature and composition fields are solved using the Crank-Nicolson scheme. Then the interface position is advanced in time to obtain a *trial* solution for the next time step, and the actual displacement of each interfacial point, as well as their temperature and solute concentration are solved iteratively [33].

The main difference between the methods of Tryggvason et al. [35, 38] and Udaykumar et al. [20, 33, 36] is the foresaid smearing of delta-functions at the interface in the former method. In the latter method, no smeared delta-functions are introduced, but the field values at the interface are used as boundary conditions for the adjacent fields. Effectively this means that cells which are cut through by the interface are divided up, and each part is "merged" with the next cell. The interface thus becomes a cell boundary. The trick is then to find the correct field update equation for these composed cells. In this paper, we closely follow the latter method by Udaykumar et al., and will focus on this update algorithm.

*3.3.3 Level set method*

Another way to describe a sharp interface is the level set method [21, 34]. This method, first introduced by Osher and Sethian [21] is similar to a phase-field model in that the solid-liquid interface is represented as the zero contour of a level set function, $\varphi(\mathbf{r},t)$, which has its own equation of motion. The movement of the interface is taken care of *implicitly* through an advection equation for φ(**r**,t), which has the physical interpretation of a (signed) distance to the interface. Unlike the phase-field model, there is no arbitrary interface width introduced in the level set method; the sharp-interface equations can be solved directly and, as a result, no asymptotic analysis is required. Discontinuous material properties can also be dealt with in a simple manner. Level set methods are particularly designed for problems in multiple space dimensions in which the topology of the evolving interface changes during the course of events, and for problems in which sharp corners and cusps are present [40]. This method was first applied to the Stefan problem on a regular grid by Gibou *et al.* [41].

The level set method has been compared to the phase-field method for dendritic growth [42, 43], and leads to the same results. However, the level set method is computationally more demanding than phase-field. This can be repaired by considering only a narrow band around the interface where the level set function is defined [40]. Moreover, a drawback of the level set method is that it is not volume preserving [22] and thus prompts the use of adaptive grid refinement [44, 45], or special enforcement of conservation laws [46]. Another important reason to apply adaptive grid refinement, is to capture small scales that would otherwise not be taken into account. This method was first applied to the Stefan problem by Chen *et al.* [47]. Yang and Udaykumar [48] describe an easy implementation of the level set method. In this respect, the immersed boundary method and the level set method are similar – they both simulate a mathematically sharp interface.

## 4 Error analysis of numerical derivatives

As mentioned above, the update scheme of Eq. (9) leads to numerical errors: the total solute mass is not conserved. It is hypothesized that the problem is caused by an error in either the (interpolated) surface concentration gradient, or the square gradient close to the interface. If the gradient – and hence the calculated growth velocity – does not exactly match the concentration reduction around the solid phase, the interface will not be displaced correctly and Eq. (2) will not be satisfied. In particular, this problem will occur when the grid is coarse as compared to the gradients that occur near the surface. For this reason, we first study how numerical errors in derivatives in 1D depend on the grid size.

Consider a Cartesian grid of spacing $h$, where the solid-liquid interface is located at position $-\varepsilon h$ (see Fig. 3). The positive direction is pointing towards the liquid phase, and the negative direction towards the solid. Without loss of generality we can choose the origin of $x$-axis at the first grid point ahead of the interface.

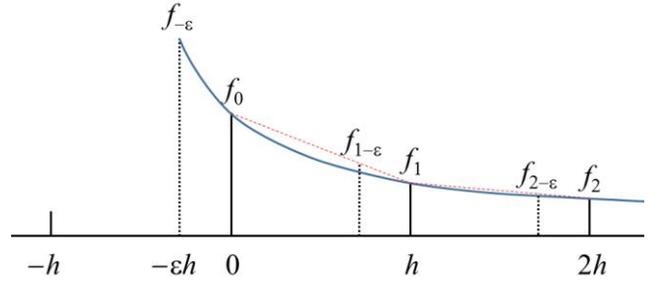

Fig. 3. Solid-liquid interface in sharp interface at grid position $-\varepsilon h$. Linear interpolations are indicated by red dotted lines.

In the method by Crank [1], a parabola is fitted through the points $x = -\varepsilon h$, $0$ and $h$, with function values $f_{-\varepsilon}$, $f_0$ and $f_1$ respectively. This leads to the following estimates (denoted by C) for the first derivative in $x = -\varepsilon h$ and for the second in $x = 0$:

$$f'_{-\varepsilon} \approx -\frac{[(1+2\varepsilon)f_{-\varepsilon} - (1+\varepsilon)^2 f_0 + \varepsilon^2 f_1]}{\varepsilon(1+\varepsilon)h}$$
$$f''_0 \approx 2\frac{[f_{-\varepsilon} - (1+\varepsilon)f_0 + \varepsilon f_1]}{\varepsilon(1+\varepsilon)h^2} \quad (C) \quad (17)$$

We test the accuracy of numerical differentiation with test function $f(x) = \exp(-\lambda(x+\varepsilon h))$. The exact derivatives ($F_{exact}$) at the interface and the next grid point are $f'_{-\varepsilon} = -\lambda$ and $f''_0 = \lambda^2 \exp(-\lambda\varepsilon h)$ respectively. The numerical values ($F_{num}$) are obtained from Eq. (17).

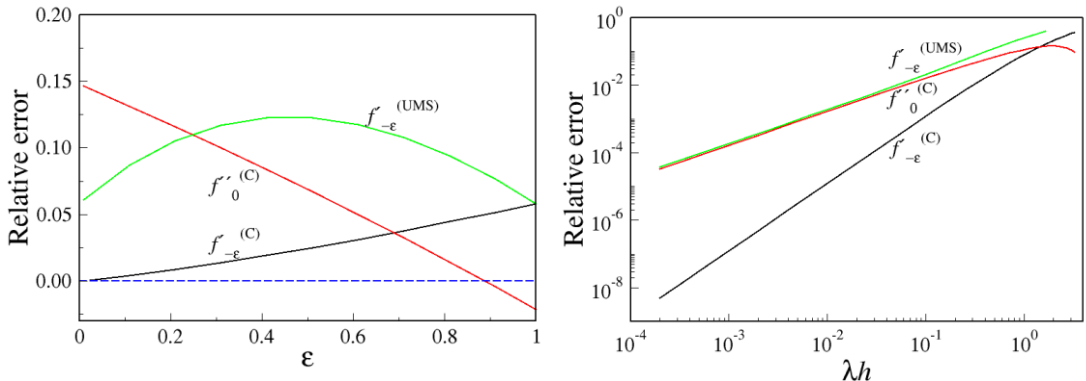

Fig. 4. Relative error in first (black) and second (red) numerical derivative using Crank's method (C) for $\lambda h = 0.5$ (left) and for $\varepsilon = 0.5$ (right), and UMS method for $f'_{-\varepsilon}$ (green). The error in Crank's method for $f'_{-\varepsilon}$ is $O(h^2)$; other errors are $O(h)$.

The relative errors, $1-F_{num}/F_{exact}$, are shown in Fig. 4. For a fixed value of $\lambda h = 0.5$ the error in the first derivative at the interface increases from 0 to 6% as $\varepsilon$ increases from 0 to 1 (black curve); and the error in the second derivative at the next grid point decreases from +15% to –2% (red curve). The scaling of the relative error with grid size is shown in the right-hand graph: the error in $f'_{-\varepsilon}$ increases $\propto h^2$, and the error in $f''_0$ increases $\propto h$. The $O(h)$ error in the square gradient at the first grid point may be a problem. Sharp concentration gradients will be a rule rather than the exception, and the power of the simulation method will be determined by its ability to represent large values of $\lambda h$. Hence, we would like to have a method that is manifestly of order $O(h^2)$.

Udaykumar et al. [20, 33] used Eq. (16) to estimate the surface gradient $f'_{-\varepsilon}$ (denoted by UMS), where the values $T_{n1}$ and $T_{n2}$ were defined at the probe positions $n1$ and $n2$. This would lead to second order accuracy if the temperature values at probe positions were exact. However, for a *bilinear* interpolation to obtain the values $T_{n1}$ and $T_{n2}$, the error in $f'_{-\varepsilon}$ is proportional to $h$. This can be seen if we check this method in one dimension. In that case, the probe positions in grid coordinates are $n1 = 1-\varepsilon$, and $n2 = 2-\varepsilon$. Interpolating the function values linearly (see Fig. 3), the first derivative at the surface becomes

$$f'_{-\varepsilon} \approx -\frac{[3f_{-\varepsilon} - 4\varepsilon f_0 + (5\varepsilon - 4)f_1 + (1-\varepsilon)f_2]}{2h} \quad (UMS) \quad (18)$$

The method used by Udaykumar to determine $f''_0$ is identical to that of Crank, Eq. (17), so this too has an accuracy of $O(h)$. Results for the error in $f'_{-\varepsilon}$, using the UMS method, are also shown in Fig. 4.

Since the second derivative determines the rate of change in the first grid point, and since the largest variations in a concentration profile will be near the interface, it is important to have this up to 2nd order in $h$. To this end we fit a 3rd order polynomial through the function values $f_{-\varepsilon} .. f_2$. Straightforward calculation now gives

$$f'_{-\varepsilon} h \approx -\frac{(2+6\varepsilon+3\varepsilon^2)}{\varepsilon(1+\varepsilon)(2+\varepsilon)} f_{-\varepsilon} + \frac{(2+3\varepsilon+\varepsilon^2)}{2\varepsilon} f_0 - \frac{(2\varepsilon+\varepsilon^2)}{(1+\varepsilon)} f_1 + \frac{(\varepsilon+\varepsilon^2)}{2(2+\varepsilon)} f_2$$

$$f''_0 h^2 \approx \frac{6}{\varepsilon(1+\varepsilon)(2+\varepsilon)} f_{-\varepsilon} + \frac{(\varepsilon-3)}{\varepsilon} f_0 + \frac{(4-2\varepsilon)}{(1+\varepsilon)} f_1 + \frac{(\varepsilon-1)}{(2+\varepsilon)} f_2$$

(19)

The expression for $f''_0$ at the first grid point near the interface, given in Eq. (19), is indeed accurate to order $O(h^2)$, see Fig. 5. As a by-product of this analysis we find an expression for $f'_{-\varepsilon}$ to $O(h^3)$. A second order determination of the square gradient near the interface requires information from the second and third grid points. Thus, if we wish to implement a Crank-Nicolson scheme, we need to invert a pentadiagonal matrix, rather than tridiagonal. An efficient code to solve a pentadiagonal system of equations is given in Ref. [31].

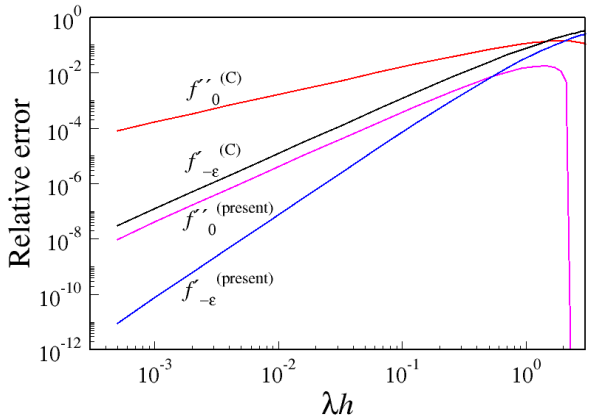

Fig. 5. Relative error in first and second numerical derivatives for Crank's method and present method for $\varepsilon = 0.5$. The present error in $f'_{-\varepsilon}$ is O($h^3$) and for $f''_0$ it is O($h^2$).

The present method for $f'_{-\varepsilon}$ is valid up to $\lambda h = 1$. At that point, the error is only 3.4%; it increases to about 14% at $\lambda h = 2$. The second derivative at the first grid point is more forgiving: the present method shows a maximum error of +1.8% at $\lambda h = 1.3$ and drops to –5% at $\lambda h = 3$. For comparison, some results are collected in Table 1.

When the decay rate is of the order of the grid spacing, $\lambda h \sim 1$, the error in the UMS method to calculate the surface gradient is some 25%. Surprisingly, the error only reduces to 20% if we replace the *linear* interpolation for the function at the probe points in Eq. (16) by *quadratic* interpolations. This limits the method to rather shallow gradients.

| $\lambda h$ | 0.2 | 0.5 | 1 | 2 |
|---|---|---|---|---|
| | Relative error [%] | | | |
| $f'_{-\varepsilon}$(C) | 0.5 | 2.5 | 7.9 | 21.0 |
| $f'_{-\varepsilon}$(UMS) | 4.4 | 12.3 | 25.3 | 45.2 |
| $f'_{-\varepsilon}$(present) | 0.05 | 0.6 | 3.4 | 13.6 |
| $f''_0$(C) | 3.1 | 6.9 | 11.3 | 14.3 |
| $f''_0$(present) | 0.14 | 0.7 | 1.5 | 0.9 |

Table 1. Relative percentage error of various methods to determine $f'_{-\varepsilon}$ and $f''_0$.

## 5    The mass conservation problem

To test the practical efficacy of the fixed grid method to simulate the Stefan problem we study the (simpler) isothermal case and check for mass conservation. For the sake of this test we set the diffusivity at $D = 1$ (independent of solute concentration); and we scale the solute concentration so that the equilibrium concentration is $C_{eq} = 1$. Grid points are chosen at $x_n = (n+½)h$, where $h = 1$, $0 \leq n \leq N–1$, and $N$ is the number of grid points. At $t = 0$ the interface is located at $s(0) = 0$, and the concentration at all grid points is given by $C(x_n) = C_n = C_0$.

At the right-hand boundary, a Neumann boundary condition is imposed, fixing the concentration *gradient* at the value of $C'(1) = 0$. This is imposed by mirroring the last point in the right-hand boundary to calculate the second derivative. To second order in $h$ we thus have: $C''(x_{N-1}) = (C_N – 2C_{N-1} + C_{N-2})/h^2 = (C_{N-2} – C_{N-1})/h^2$. At the left-hand boundary we impose the Stefan condition $ds/dt = -C'(s,t) = -C_s'(t)$, together with the equilibrium condition $C(s,t) = 1$. For each value of $s$ we obtain the *minimum* grid number $k$ for which the diffusion equation applies, and we obtain the distance ε between $s$ and $x_k$ in grid units, as

$$k = \text{int}(s/h + \tfrac{1}{2})$$
$$\varepsilon = k + \tfrac{1}{2} - s/h \quad (20)$$

Thus, we arrive at the following set of equations:

$$\begin{cases} s(t+\delta t) - s(t) = -\tfrac{1}{2}\delta t \cdot (C'_s(t) + C'_s(t+\delta t))/C_s \\ C_k(t+\delta t) - \tfrac{1}{2}\delta t \cdot C''_k(t+\delta t) = C_k(t) + \tfrac{1}{2}\delta t \cdot C''_k(t) \\ C_n(t+\delta t) - \tfrac{1}{2}\delta t \cdot (C_{n+1}(t+\delta t) - 2C_n(t+\delta t) + C_{n-1}(t+\delta t))/h^2 = C_n(t) + \tfrac{1}{2}\delta t \cdot C''_n(t) \\ C_{N-1}(t+\delta t) - \tfrac{1}{2}\delta t \cdot (C_{N-2}(t+\delta t) - C_{N-1}(t+\delta t))/h^2 = C_{N-1}(t) + \tfrac{1}{2}\delta t \cdot C''_{N-1}(t) \end{cases} \quad (21)$$

In this set of equations $k < n < N-1$. It should be noted that $k$ is initially $k = 0$. However, when $s(t)$ crosses point $x_k$ between $t$ and $t+\delta t$, we need to increase $k$ by 1 and solve the set of equations again for fewer grid points. Moreover, hidden in the first two lines of Eq. (21) is an *implicit* dependence on $s(t+\delta t)$, as the derivatives mentioned there depend in a non-linear way upon $\varepsilon(t)$ and $\varepsilon(t+\delta t)$ (see Eq. (17) and Eq. (19)). For these reasons the system of equations needs to be solved in an iterative way, alternating between a Crank-Nicolson solution for the concentration field at $t+\delta t$ with a fixed trial value for the interface position, and a solution method to obtain $s(t+\delta t)$ for a fixed concentration field.

It is important to note that $C''_k(t)$, as appearing at the right-hand side of Eq. (21), can be either a special second derivative, as given by Eq. (19), or an ordinary second derivative on a regular grid. If $s(t+\delta t)$ passes a grid point, the $k$-value is increased, hence grid point $k$ of the *old* concentration field $C_k(t)$ is no longer adjacent to the ice front. Therefore, the second derivative of the old concentration field is stored separately before we enter the iterative scheme to determine $s(t+\delta t)$. Note further that $C''_k(t+\delta t)$ depends not only on $\varepsilon(t+\delta t)$, but also on the concentration at the interface (see Eq. (19)). This is a (fixed) input term, and is therefore taken to the right-hand side of the equation for the solution procedure.

To find $s(t+\delta t)$ we use the following scheme:

Step 1: estimate new front position from

$$s^0(t+\delta t) = s(t) - \delta t \cdot C'_s(t)/C_s \quad (22)$$

and set counter $j = 1$.

Step 2: determine $k$ and ε for the new front position from Eq. (20).

Step 3: solve $C(s,t+\delta t)$ from the last three lines of Eq. (21) for a *fixed* value of $s(t+\delta t)$

Step 4: determine new front position from

$$s^j(t+\delta t) = s(t) - \tfrac{1}{2}\delta t \cdot (C'_s(t) + C'_{s^{j-1}}(t+\delta t))/C_s \quad (23)$$

and increase counter $j = j+1$.

Step 5: return to Step 2 if $j \leq 4$.

Step 5 could be replaced by a convergence criterion, but it is found that four iterations suffices to obtain six decimal places accuracy in the new position, even for a time large time step like $\delta t = 0.5h^2/D$. For simplicity a fixed number of iterations is taken. After four iterations the new concentration field is copied onto the old field, and time is increased by $\delta t$. The square gradient of the concentration field is then stored for input to the next time step, and the next time step is taken.

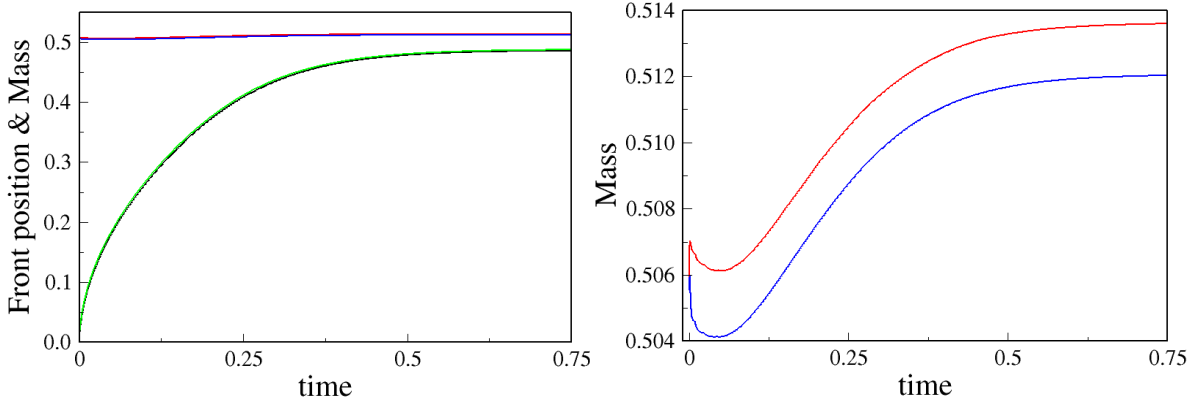

Fig. 6. Front position and mass for isothermal simulation using Crank's equation for the surface gradient (black and red) and using Eq. (19) for the surface gradient (green and blue).

To test the accuracy of this scheme, simulations are done for small systems of $N = 20$ grid points. An error in mass conservation will be more pronounced in small systems then in large systems because the error is a surface effect. In the presentation below we rescale the system size to length $L = 1$. We start at each grid point with the initial concentration $C_0 = 0.5$, and we maintain the equilibrium concentration $C_s = 1$. In these units, the diffusion constant is $D = h^2 = 1/N^2$, hence we rescale time to $t^* = Dt/L^2$. Furthermore, the total mass in the system is $M = \int C_0 dx = 0.5$, hence the front position should converge to $s(t) = 0.5$ for $t^* \to \infty$.

The front position and total mass in the system are shown in Fig. 6a. The green and black curves give the front position, calculated from two approximations for the concentration gradient at the ice front. The black curve is for Crank's approximation (Eq. (17)), and the green curve is for the present 3rd order approximation (Eq. (19)). The red and blue curves respectively give the total mass for the two simulations. The mass is calculated by integrating piecewise parabola interpolations. Note that we start with too high mass, because the first grid point at $x = h/2$ has concentration $C = C_0 = 0.5$, but at the interface at $x = 0$ we have $C = C_s = 1$. Hence the mass integral is roughly $M \approx 0.5 + 0.25 \cdot h/2 = 0.506$. Fig. 6b shows that this value is not conserved, although the maximum deviation from the mean value is of the order of 0.5%, and the variation between maximum and minimum of the curve is about 1.5%. Surprisingly, using a third order polynomial to estimate the surface gradient (blue curve in Fig. 6b) has hardly any advantage over a parabola. Probably this is because the error made in the second derivative is of order $h^2$, which will dominate the error in the surface gradient ($\propto h^3$).

To stabilize the total solute mass in the simulation, a correction method is sought for. Since the gradient is not determined exactly, this is a likely source of errors. And if the front position is not updated in balance with the concentration increase the integrated mass will start to shift in the course of time. Therefore, the position update scheme is corrected to

$$s(t + \delta t) = s(t) - Df\delta t \cdot \nabla C / C_s \qquad (24)$$

where $D$ is the diffusivity, $\nabla C$ is the mean concentration gradient over the time interval $\delta t$ as determined by the interpolation procedure, and $f \approx 1$ is a gradient correction factor. Let $m$ be the mass determined by numerical integration. Using piecewise linear interpolation (for vanishing slope at the right wall) we obtain

$$m/h = \tfrac{1}{2}\varepsilon(C_s + C_k) + \tfrac{1}{2}\sum_{n=k}^{N-2}(C_n + C_{n+1}) + \tfrac{1}{2}C_{N-1} \quad (25)$$

Now let $M$ be the exact mass, which should be conserved. If in a previous time step mass $m$ deviates from $M$ we need to shift the front position by an amount $\delta s$ such that $m - C_s \delta s = M$, or

$$\delta s = \frac{m - M}{C_s} = -D(f-1)\delta t \cdot \frac{\nabla C}{C_s} \qquad (26)$$

The gradient correction factor to force mass conservation is thus found as

$$f = 1 + \frac{M - m}{D\delta t \nabla C} \qquad (27)$$

Note that the gradient correction factor is dimensionless, hence the same equation applies whether we use the grid size as unit of length, or any other choice for the unit of length.

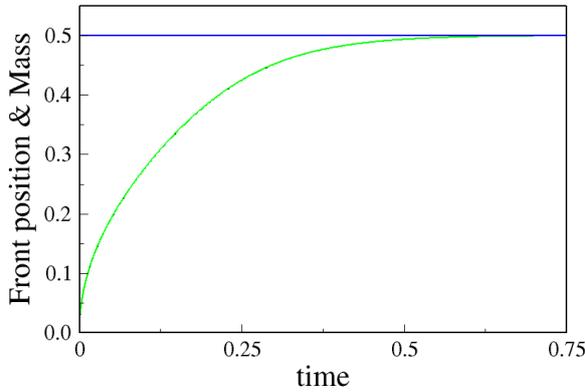
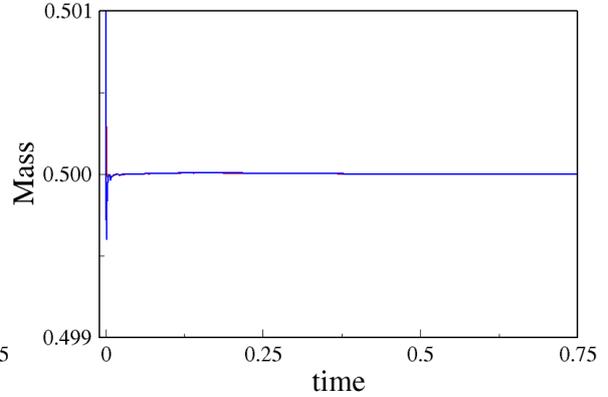

Fig. 7. Front position and mass for isothermal simulation using Crank's equation for the surface gradient (black and red) and using Eq. (19) for the surface gradient (green and blue), imposing correction factor $f$ (Eq. (27)) at every time step.

When the gradient correction factor is calculated from the field gradient in the last time step (Step 1, Eq. (22)) and subsequently applied in Step 4 above (Eq. (23)) to obtain the next front position, we find the results shown in Fig. 7. Whereas in Fig. 6 a small difference can be seen between the front positions obtained from the results from 2nd and 3rd order polynomials to calculate the surface gradient, in Fig. 7 the two curves superimpose exactly. Moreover, they converge exactly to the correct value $s(t) = 0.5$ for $t^* \to \infty$; and throughout the simulation the total solute mass equals $m = 0.5$. Deviations typically occur in the sixth decimal place. One might expect the scheme to oscillate, but it does not. An enlargement of the total solute mass is shown in Fig. 7b. this shows the stability of the correction scheme, only during the first few time steps significant deviations occur because we start with the wrong mass $m \approx 0.506$, because of the finite grid size.

To test the efficacy of this scheme, the grid size was varied in a finite size scaling analysis from $h = 1/160$ to $1/5$. We find the simulation results shown in Fig. 8. The left picture shows the simulation without correction factor, and the right-hand graph shows the results with correction factor. The correction factor does correct the total mass content, and hence the final grid position, but the dynamics with correction clearly shows a dependence on grid size. Without correction, all curves superimpose up to $t^* \approx 0.22$ and then start to deviate, but with correction factor a maximum deviation occurs near $t^* \approx 0.26$, and then the curves converge again. Hence, although the correction factor solves the mass conservation problem, it introduces another problem.

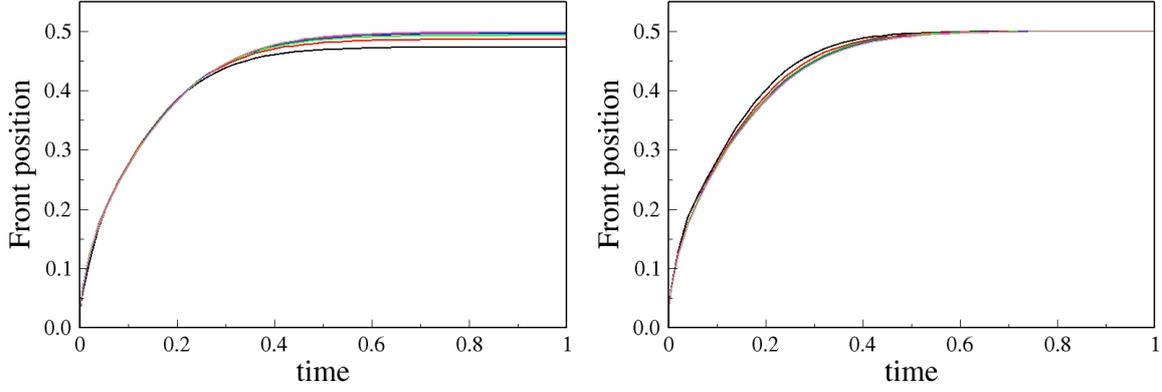

Fig. 8. Simulation for various grid sizes without correction factor (left) and with correction (right). Without correction mass is not conserved, but with correction dynamics depends on grid size.

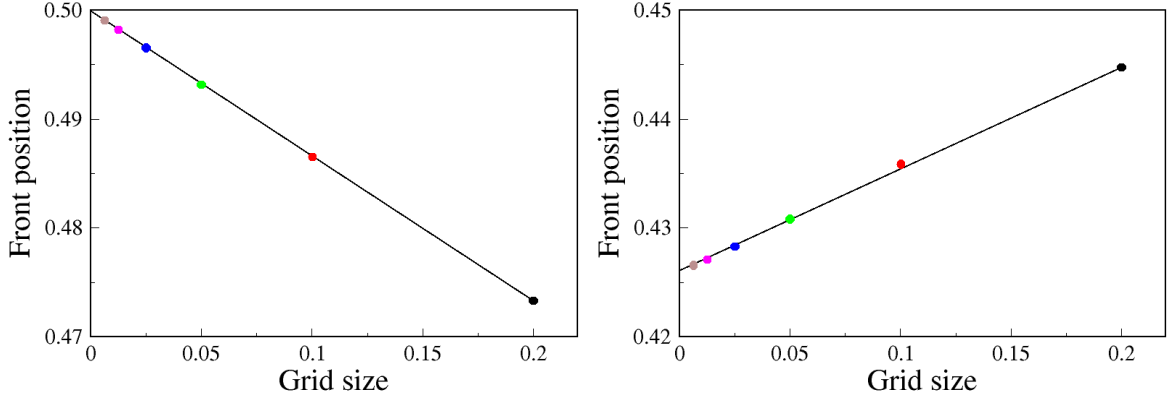

Fig. 9. Front position as function of grid size at $t^* = Dt/L^2 = 1$ without correction factor (left), and at $t^* = Dt/L^2 = 0.26$ with correction (right). In both cases the error is linear in the grid size.

If we plot the front position as function of grid size, we find that both with and without correction an error proportional to the grid size $h$ is made, see Fig. 9. This is unfortunate, as it implies that we would need a very fine mesh for accurate simulations. This in turn means that generalisation to 2D or 3D will be computationally intensive.

To summarize, even though we have included expressions for the first and second order derivatives near the interface that are accurate to second order, we still find a first order error in the update scheme. Hence, we conclude that the initial hypothesis, that errors are caused by inaccurate numerical derivatives, is incorrect. Thus, we need to search for a new update scheme that is inherently mass conserving, and is accurate to order $h^2$. This is the subject of the next section.

## 6 A mass conserving second order update scheme

To conserve mass, we make a choice for the density distribution function. We maintain grid points at half-integer lattice positions, $x_n/h = (n+½)$, and we *define* the density distribution at intermediate points by linear interpolations. An example is shown in Fig. 10. Each cell $n$ runs from $n < x/h < n+1$; and the mass flux from cell $n-1$ to cell $n$, $J_n$, is located at $x/h = n$, see Fig. 10. At the right-hand boundary $x = L = Nh$ we impose an insulating boundary condition. This can be imposed by mirroring the concentration field in $x = L$, which implies that the concentration field runs horizontal between $L-h/2$ and $L$. For this concentration field the total mass in *lattice units* is *exactly* given by

$$m = \tfrac{1}{2}\varepsilon(C_s + C_k) + \tfrac{1}{2}\sum_{n=k}^{N-2}(C_n + C_{n+1}) + \tfrac{1}{2}C_{N-1}$$
$$= \sum_{n=k+1}^{N-1} C_n + \tfrac{1}{2}\varepsilon C_s + \tfrac{1}{2}(1+\varepsilon)C_k \quad (28)$$

This approximates the mass of an *arbitrary* concentration profile to order $h^2$.

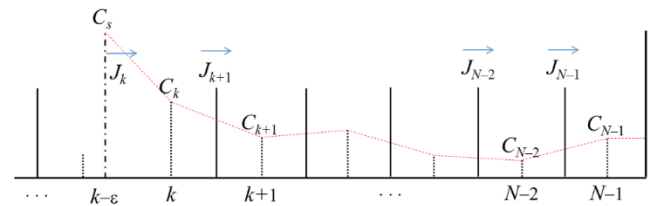

Fig. 10. Sharp solid-liquid interface at grid position $k-\varepsilon$. Linear interpolations are indicated by red dotted lines. Concentration fluxes and diffusivities are defined at cell boundaries.

Using Fick's law the mass fluxes between successive cells is (using grid units, $h = 1$)

$$\begin{cases} J_k = -D_s \nabla C_s \\ J_n = D_n(C_{n-1} - C_n) \quad (n > k) \end{cases} \quad (29)$$

where $D_n$ is the *local* diffusivity, which is defined at the location of $J_n$, and may depend on the temperatures and solute concentrations of the neighbouring cells: $D_n = D_n((C_{n-1}+C_n)/2, (T_{n-1}+T_n)/2)$. $D_s$ is the diffusivity at the growing surface. No mass is passing through the right boundary, therefore the rate of change of the concentration in the last bin is given by

$$\dot{C}_{N-1} = J_{N-1} = D_{N-1}(C_{N-2} - C_{N-1}) \quad (30)$$

For all cells that are not adjacent to the ice front we have

$$\dot{C}_n = J_n - J_{n+1} \quad (k < n < N-1) \quad (31)$$

Therefore we have

$$\sum_{n=k+1}^{N-1} \dot{C}_n = J_{N-1} + (-J_{N-1} + J_{N-2}) + \\ \ldots + (-J_{k+2} + J_{k+1}) = J_{k+1} \quad (32)$$

Hence the time derivative of the total mass is

$$\dot{m} = J_{k+1} + \tfrac{1}{2}\dot{\varepsilon}(C_s + C_k) + \tfrac{1}{2}\varepsilon\dot{C}_s + \tfrac{1}{2}(1+\varepsilon)\dot{C}_k \quad (33)$$

Using the Stefan condition ($\dot{\varepsilon} = -J_k/C_s$) and imposing mass conservation we obtain the time derivative of $C_k$ as

$$\dot{C}_k = \frac{J_k(1 + C_k/C_s) - 2J_{k+1} - \varepsilon\dot{C}_s}{(1+\varepsilon)} \quad (34)$$

It should be noted that the right-hand side of Eq. (34) is *not* equal to the second derivative at the first grid point; there are corrections to O($h$). This is the reason why the update scheme of the previous section (Eq. (21)) is only correct to O($h$). Reversely, if we base the time derivative on a square gradient that is correct to O($h^2$), mass conservation *must be violated* to O($h$). This explains the violation of mass conservation shown in Fig. 8a and Fig. 9a. If on the other hand, we use an O($h^2$) square gradient for $C_k$ and force mass conservation by changing the motion of the front, this motion must deviate from the correct Stefan condition, resulting in an O($h$) error in the front position. This explains the O($h$) deviation from the correct front evolution, shown in Fig. 8b and Fig. 9b.

In general, we can use *any* approximation for the flux $J_k$ at the surface, as long as *the same* approximation is used for the growth rate of the concentration in the first lattice point $C_k$ (Eq. (34)) *and* for the motion of the interface through the Stefan condition. Two approximations for the gradient at the surface as given in Eq. (17) and Eq. (19) are reproduced here:

$$\nabla C_s = -\frac{(1+2\varepsilon)}{\varepsilon(1+\varepsilon)}C_s + \frac{(1+\varepsilon)}{\varepsilon}C_k - \frac{\varepsilon}{(1+\varepsilon)}C_{k+1} \qquad [O(h^2)]$$
$$\nabla C_s = -\frac{(2+6\varepsilon+3\varepsilon^2)}{\varepsilon(1+\varepsilon)(2+\varepsilon)}C_s + \frac{(2+3\varepsilon+\varepsilon^2)}{2\varepsilon}C_k - \frac{(2\varepsilon+\varepsilon^2)}{(1+\varepsilon)}C_{k+1} + \frac{(\varepsilon+\varepsilon^2)}{2(2+\varepsilon)}C_{k+2} \qquad [O(h^3)] \quad (35)$$

Thus, we may write the surface gradient in general as $\nabla C_s = aC_s + b_0 C_k + b_1 C_{k+1} + b_2 C_{k+2}$; the coefficients can be read off from Eq. (35).

To arrive at a forward integration scheme for *finite time steps* we start with the mass balance

$$\sum_{n=k+1}^{N-1} C_n + \tfrac{1}{2}\varepsilon C_s + \tfrac{1}{2}(1+\varepsilon)C_k = \sum_{n=k'+1}^{N-1} C'_n \\ + \tfrac{1}{2}\varepsilon' C'_s + \tfrac{1}{2}(1+\varepsilon')C'_{k'} \quad (36)$$

where the primed quantities are for the new time step. For simplicity, we consider an explicit update scheme; generalisation to a Crank-Nicolson scheme is straightforward but tedious in programming. First, we note that the sum over $C'_n$ at the right-hand side follows from Eq. (32) as

$$\sum_{n=k'+1}^{N-1} C'_n = \sum_{n=k'+1}^{N-1} C_n + J_{k'+1}\delta t \quad (37)$$

The index $k'$ denotes the number of the grid point adjacent to the front, in the next time step. Now two cases may occur, where $k' = k$ or $k' = k+1$. In the former case the front does not cross a grid point, in the latter case is does.

We start with the first case, where $k' = k$. In a forward integration scheme the distance from front to the first grid point evolves according to the Stefan condition as

$$\varepsilon' = \varepsilon - \delta t \cdot J_k/C_s \quad (38)$$

where $J_k = -D_s \nabla C_s$, and $D_s$ is the diffusivity at the surface. Substitution of Eq. (37–38) into Eq. (36), and using $k' = k$, we find the first layer concentration in the next time step as

$$C'_k = \frac{(1+\varepsilon)C_k + \varepsilon(C_s - C'_s) + (J_k C'_s/C_s - 2J_{k+1})\delta t}{(1+\varepsilon')} \\ = C_k + \frac{J_k(1 + C_k/C_s) - 2J_{k+1} - \varepsilon'\dot{C}_s}{(1+\varepsilon')}\delta t \quad (39)$$

Note that in the limit of continuous time this is identical to the time derivative obtained in Eq. (34), but if we account for a finite time step we need to replace $\varepsilon$ by $\varepsilon'$ at the right-hand side to obtain *exact* mass conservation. The time derivative of the surface concentration used here is the forward time derivative, $\dot{C}_s = (C_s(t+\delta t) - C_s(t))/\delta t$.

We now consider the second case, where $k' = k+1$. In this case, the ice front crosses a grid point, hence the distance from front to the new first grid point changes to

$$\varepsilon' = \varepsilon - \delta t \cdot J_k/C_s + 1 \quad (40)$$

Because the sum over $C_n$ at the left-hand side of Eq. (36) starts at $k+1$, and at the right-hand side it starts at $k'+1 = k+2$, a term $C_{k+1}$ is left over at the left-hand side which does not cancel out. Moreover, when $\varepsilon'$ is substituted into Eq. (36), the +1 term in Eq. (40) leads to an extra $C'_s$ term at the right-hand side. Collecting terms, we find

$$C'_{k+1} = \frac{2C_{k+1} - C'_s + (1+\varepsilon)C_k + \varepsilon(C_s - C'_s) + (J_k C'_s/C_s - 2J_{k+2})\delta t}{(1+\varepsilon')} \quad (41)$$

$$= C_{k+1} + \frac{J_k(1+C_k/C_s) - 2J_{k+2} - \varepsilon'\dot{C}_s}{(1+\varepsilon')}\delta t + \frac{(1-\varepsilon')C_{k+1} + \varepsilon'C_k - C_s}{(1+\varepsilon')}$$

Thus, we find the same term as time derivative for $C_{k+1}$ as for $C_k$ for the case where no grid point was passed, albeit that a $J_{k+1}$ term is replaced by $J_{k+2}$. Remarkably, an extra term appears that cannot be identified easily as a time derivative, but it can be interpreted as a concentration gradient. To generalise this evolution algorithm to a Crank-Nicolson scheme we should replace Eq. (38) and Eq. (40) by $\varepsilon' = \varepsilon - \tfrac{1}{2}\delta t \cdot J_k/C_s - \tfrac{1}{2}\delta t \cdot J'_k/C'_s$ and $\varepsilon' = \varepsilon - \tfrac{1}{2}\delta t \cdot J_k/C_s - \tfrac{1}{2}\delta t \cdot J'_{k+1}/C'_{s+1}$ respectively, and solve for the primed variables.

Mass conservation was checked for systems of $N = L/h = 5$ to 80 grid points, using a fixed time step of $D\delta t/h^2 = 0.0025$. The systems were integrated to time $Dt/L^2 = 1$. The initial and final mass integral varied at most by a relative error of $10^{-14}$, i.e. it is constant to the machine precision. This is far superior to the results shown in Fig. 6 and Fig. 7.

To check the scaling to grid size, the grid spacing was varied from $h = 0.2$ down to $h = 0.0125$. To ascertain that all systems had equal mass from the start, the initial value of the front position was varied. Solving $\varepsilon$ from $\tfrac{1}{2}\varepsilon C_s + \tfrac{1}{2}(1+\varepsilon)C_0 = C_0$ (the mass in the first grid cell), we find the initial front position as

$$s_0 = \tfrac{1}{2}\frac{C_s - C_0}{C_s + C_0} \quad (42)$$

One may argue that simulations for different grid size start at a different times $t_0$, as coarser grids have a head start. As the front position initially evolves as $s(t) = (\lambda t)^{1/2}$ (see Appendix), the time offset is obtained as $t_0 = s_0^2/\lambda$, where $\lambda$ is obtained as the slope in a plot of $s^2$ as function of $t$. After correcting for this offset we find the results shown in Fig. 11.

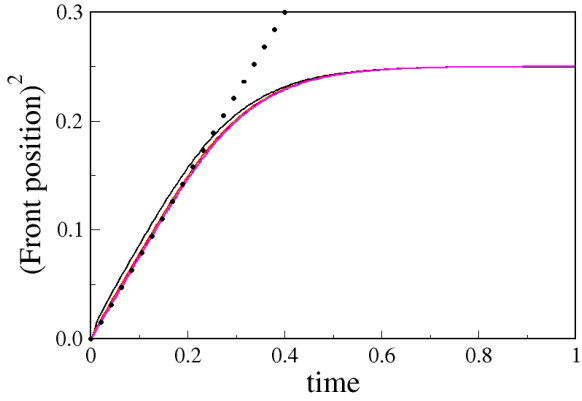 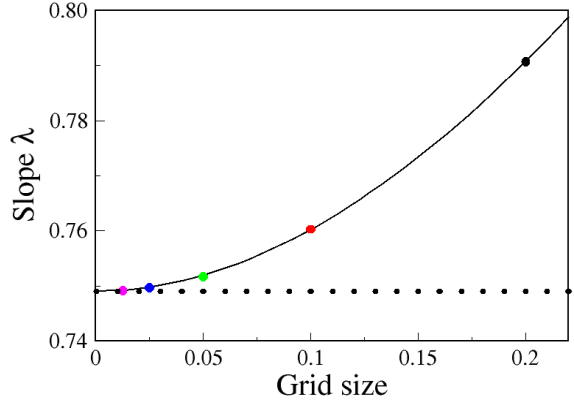

Fig. 11. Front position for various grid sizes after rescaling the initial front position (left) for grid size $h = 0.2$, 0.1, 0.05, 0.025 and 0.0125; and the initial slope of $s^2$ as function of grid size (right). Dotted lines indicate analytic limiting law.

Fig. 11a shows excellent scaling. Only the data for the largest grid size (black line) deviates visibly from the data obtained from smaller grid size. The black dotted lines gives the analytical initial behaviour $s^2 = \lambda t$, where $\lambda$ is obtained from

$$\Delta = (1 - C_0/C_s) = \tfrac{1}{2}\sqrt{\pi\lambda} \cdot \mathrm{erfc}\left(\tfrac{1}{2}\sqrt{\lambda}\right) \cdot \exp\left(\tfrac{1}{4}\lambda\right) \quad (43)$$

A derivation is given in the Appendix. For $\Delta = 0.5$ we find $\lambda_{exact} = 0.749096$.

To obtain a measure of the numerical error, the initial slopes of the curves in Fig. 11a have been determined for $0.015 < Dt/L^2 < 0.1$. The results are shown as function of grid size in Fig. 11b. The black dotted line again gives the analytical result $\lambda_{exact} = 0.749096$. The numerical results deviate from the exact result as $\lambda_{num} - \lambda_{exact} \approx 1.2h^2 - 0.8h^3$, i.e. the error is indeed of second order in the grid size. Note that the *relative* error is quite small, only 1.5% for $h = 0.1$ (ten grid points), and 0.35% for $h = 0.05$. Even for 5 grid points in the simulated space we have a modest error of 5.5%. This opens the door to large scale simulations with modest computer power.

## 7 Summary and conclusions

A brief review of the Stefan problem and the main numerical solution methods is given. We concentrate on the problem of solidification from solution. For this problem, calculations in 2D or 3D are most practically done on a regular grid, where the solid front moves relative to the grid. In the phase-field method the interface is spread out over several grid points. Moreover, for mixtures the width of the interface in grid units should be smaller than the range of the solute diffusion field. This implies that many grid points are needed for simulations of realistic complexity, which is computationally intensive.

Alternative methods are level set and explicit front tracking methods, of which the latter are conceptually easier. Physical boundary conditions at the moving interface can be applied straightforwardly. The motion of the interface depends on the temperature and solute gradients at the interface, and the evolution of the concentration and temperature fields away from the interface are determined by square gradients. In current implementations, these are determined to an error that is linear in the grid size. This may be a problem.

Sharp concentration gradients will be a rule rather than the exception, and the power of a simulation method is determined by its ability to represent large field variations with few grid points. Therefore, higher order interpolations are used to obtain the gradient at, and square gradient near the interface, to second order in the grid size.

It is surprisingly found that, even with such second order precision in the field derivatives, mass is conserved only up to first order in the grid size. This is a problem, because it implies that still many grid points are needed in 2D and 3D calculations to obtain acceptable accuracy, leading to large computational costs.

Therefore, a new method is developed from the leading principle of local mass conservation. In this method, the motion of the interface is accurate to *second order* in the grid size. In this method, the time derivative for the field point adjacent to an interface is *not equal to the square gradient* in that point, but contains corrections to first order in the grid size. An update scheme for finite time steps is derived. When the interface crosses a grid point during a time step, the update scheme for this first grid point contains an *additional* term which can be interpreted as a gradient term, rather than a square gradient.

When applied to 1D simulations, we find manifest mass conservation to 14 decimal places. To test the method, we study an isothermal problem with a supersaturation of 50%. When only 5 grid points are used in the interval, we find the growth velocity of the interface accurate to 5.5% as compared to the exact result, and for 10 points the result is accurate to 1.5%. It is anticipated that the method can be generalized to 2D and 3D simulations with relative ease. This opens the door to large scale simulations of the Stefan problem with modest computer power.

**Appendix -** *Initial growth law for isothermal conditions*

We consider a half space where the sucrose concentration at $t = 0$ is given by $C_0$ for $x > 0$, and has a fixed value $C_s$ at the surface. The concentration profile follows $\partial C/\partial t = \partial^2 C/\partial x^2$, and the front moves as $ds/dt = -\nabla C/C_s$. We now use a scaling ansatz [1] where the concentration profile only depends on the combination $\xi = x/\sqrt{t}$, hence $C(x,t) = u(x/\sqrt{t}) = u(\xi)$. This planar case is similar to the classical Frank sphere solution [1], and is given here for completeness. Using $\partial/\partial x = t^{-1/2} \partial/\partial \xi$ and $\partial/\partial t = -\tfrac{1}{2} t^{-3/2} \partial/\partial \xi$, the diffusion equation becomes $u'' + \tfrac{1}{2}\xi u' = 0$, which is solved directly as $u' = \exp(-\tfrac{1}{4} \xi^2 + constant)$.

Integrating again gives the general solution $C(x,t) = u(x/\sqrt{t}) = A \operatorname{erf}(\tfrac{1}{2} xt^{-1/2})+B$. Applying boundary conditions $C(s(t)) = C_s$ and $C(\infty) = C_0$, we solve $A$ and $B$ from the system $C_s = A \operatorname{erf}(\tfrac{1}{2} st^{-1/2})+B$ and $C_0 = A+B$. Because $C_s$ is a constant, $s(t)$ must be proportional to $t^{1/2}$ to cancel the factor $t^{-1/2}$ in the error function of the general solution. Thus, we only have a time-independent boundary condition at the moving interface when the interface moves with the scaling law

$$s = (\lambda t)^{1/2}$$

where $\lambda$ is a constant. In other words, the time-independence of the boundary condition at the surface forces the surface motion to a square root time behaviour. Solving for $A$ and $B$ we obtain the solution

$$C(x,t) = C_0 + (C_s - C_0)\operatorname{erfc}(\tfrac{1}{2} xt^{-1/2})/\operatorname{erfc}(\tfrac{1}{2} \lambda^{1/2})$$

Now we apply the Stefan condition to find $\lambda$, and obtain

$$\dot{s} = -\frac{\nabla C(s,t)}{C_s} = \frac{(C_s - C_0)}{C_s} \cdot \frac{1}{\sqrt{\pi t}} \cdot \frac{\exp(-\tfrac{1}{4} s^2/t)}{\operatorname{erfc}(\tfrac{1}{2}\lambda^{1/2})} = \tfrac{1}{2}\sqrt{\frac{\lambda}{t}}$$

where the final step follows from the time derivative of the above square root time behaviour of the moving front. We finally obtain the implicit relation between the rate constant $\lambda$ and the supersaturation:

$$\Delta = (1 - C_0/C_s) = \tfrac{1}{2}\sqrt{\pi\lambda} \cdot \operatorname{erfc}(\tfrac{1}{2}\sqrt{\lambda}) \cdot \exp(\tfrac{1}{4}\lambda) \quad (A1)$$

To obtain the inverse function, we first study the limits for small and large values of $\lambda$. For small $\lambda$ we obviously have $\lambda \approx 4\Delta^2/\pi$, and in the limit of large $\lambda$ we find the asymptotic expansion $\lambda \approx 1-2/\Delta$. Hence a reasonable first approximation for $\lambda$ is:

$$\lambda \approx \frac{4\Delta^2/\pi + (2 - 4/\pi)\Delta^4}{(1-\Delta)} \quad (A2)$$

This approximation satisfies the limits for small and large $\lambda$, hence it forms a good starting point for a numerical solution for the inverse of Eq. (A1). Using the Newton-Raphson method, a next approximation to $\lambda$ is obtained as

$$\lambda' = \lambda - \frac{4(\Delta - \Delta_0)}{(\Delta + 2\Delta/\lambda - 1)} \quad (A3)$$

where $\Delta$ is the function given in Eq. (A1), and $\Delta_0$ is the desired value of the supersaturation for which we want to calculate the corresponding slope $\lambda$. Applying Eq. (A3) iteratively, a solution to 14 decimal places is typically obtained in three iterations, when the estimate of Eq. (A2) is used as a starting point.

**References**

[1] J. Crank, The Mathematics of Diffusion, second ed., Clarendon press, Oxford, 1975, pp. 286-325.
[2] J. Stefan, Ueber die Theorie der Eisbildung, insbesondere über die Eisbildung im Polarmeere, Annalen der Physik 278 (1891) 269-286.
[3] C. Körber and G. Rau, Ice Crystal Growth in Aqueous Solutions, in: D.E. Pegg, A.M. Karow Jr., (eds.), The Biophysics of Organ Cryopreservation, Plenum Press, New York, 1987, pp. 173-199.
[4] K. Wollhöver. Ch. Körber, M.W. Scheiwe and U. Hartmann, Unidirectional freezing of binary aqueous solutions: an analysis of transient diffusion of heat and mass, Int. J. Heat Mass Transfer 28 (1985) 761-769.
[5] M.F. Butler, Instability Formation and Directional Dendritic Growth of Ice Studied by Optical Interferometry, Cryst. Growth Des. 1 (2001) 213-223.
[6] M.F. Butler, Growth of Solutal Ice Dendrites Studied by Optical Interferometry, Cryst. Growth Des. 2 (2002) 59-66.

[7] M.A. Martorano and J.D.T. Capocchi, Dendrite structure control in directionally solidified bronze castings, Int. J. Cast Metals Res. 13 (2000) 49-57.

[8] R.A. Martinez, A. Karma and M.C. Flemings, Spheroidal Particle Stability in Semisolid Processing, Metall. Mater. Trans. 37A (2006) 2807-2815.

[9] J. Wannasin, R.A. Martinez and M.C. Flemings, Grain refinement of an aluminum alloy by introducing gas bubbles during solidification, Scripta Mater. 55 (2006) 115-118.

[10] J.S. Langer, Instabilities and pattern formation in crystal growth, Rev. Mod. Phys. 52(1980) 1-28.

[11] M. Asta, C. Beckermann, A. Karma, W. Kurz, R. Napolitano, M. Plapp, G. Purdy, M. Rappaz, and R. Trivedi, Solidification microstructures and solid-state parallels: Recent developments, future directions, Acta Materialia 57 (2009) 941-971.

[12] A.A. Shibkov, Y.I. Golovin, M.A. Zheltov, A.A. Korolev and A.A. Leonov, Morphology diagram of nonequilibrium patterns of ice crystals growing in supercooled water, Physica A 319, (2003) 65-79.

[13] W.W. Mullins and R.F. Sekerka, Stability of a Planar Interface During Solidification of a Dilute Binary Alloy, J. App. Phys. 35 (2) (1964) 444-451.

[14] W.W. Mullins and R.F. Sekerka, Morphological Stability of a Particle Growing by Diffusion or Heat Flow, J. App. Phys. 34 (2) (1963) 323-329.

[15] R. Trivedi and W. Kurz, Morphological Stability of a Planar Interface Under Rapid Solidification Conditions, Acta Metall. 34 (8) (1986) 1663-1670.

[16] J.S. Langer, H. Müller-Krumbhaar, Theory of Dendritic Growth – I. Elements of a Stability Analysis, Acta Metall. 26 (1978) 1681-1687.

[17] J.S. Langer and H. Müller-Krumbhaar, Stability Effects in Dendritic Crystal Growth, J. Cryst. Growth 42 (1977) 11-14.

[18] J.S. Langer, R.F. Sekerka and T. Fujioka, Evidence for a Universal Law of Dendritic Growth Rates, J. Cryst. Growth 44 (1978) 414-418.

[19] A. Karma and W.J. Rappel, Quantitative phase-field modeling of dendritic growth in two and three dimensions, Phys. Rev. E 57 (4) (1998) 4323-4349.

[20] H.S. Udaykumar, R. Mittal and W. Shyy, Computation of Solid–Liquid Phase Fronts in the Sharp Interface Limit on Fixed Grids, J. Comput. Phys. 153 (1999) 535-574.

[21] S. Osher and J.A. Sethian, Fronts propagating with curvature dependent speed: algorithms based on Hamilton-Jacobi formulations, J. Comput. Phys. 79 (1988) 12-49.

[22] A.K. Tornberg and B. Enhquist, A finite element based level set method for multiphase flow applications, Comput. Visual. Sci. 3 (2000) 93-101.

[23] T. van Westen and R.D. Groot, Predicting the Kinetics of Ice Recrystallization in Aqueous Sugar Solutions, Cryst. Growth Des. 18 (2018) 2405–2416.

[24] S. Marella, S. Krishnan, H. Liu and H.S. Udaykumar, Sharp interface Cartesian grid method I: An easily implemented technique for 3D moving boundary computations, J. Comput. Phys. 210 (2005) 1-31.

[25] X. Zheng, H. Babaee, S. Dong, C. Chryssostomidis and G.E. Karniadakis, A phase-field method for 3D simulation of two-phase heat transfer, Int. J. Heat Mass Trans. 82 (2015) 282-298.

[26] R.B. Bird, W.E. Stewart and E.N. Lightfoot, Transport Phenomena, second ed., Wiley, New York, 2002, pp. 83-84.

[27] R. Mittal, H. Dong, M. Bozkurttas, F.M. Najjar, A. Vargas and A. von Loebbecke, A versatile sharp interface immersed boundary method for incompressible flows with complex boundaries, J. Comput. Phys. 227 (2008) 4825-4852.

[28] R. Vichnevetsky, A new stable computing method for the serial hybrid computer integration of partial differential equations, in: AFIPS Conference Proceedings, Vol. 32: 1968 Spring Joint Computer Conference, Thompson, New Jersey, 1968, pp. 143-150.

[29] T.Y. Hou, J.S. Lowengrub, and M.J. Shelley, Removing the stiffness from interfacial flows with surface tension, J. Comput. Phys. 114 (1994) 312-338.

[30] J. Crank and P. Nicolson, A practical method for numerical evaluation of solutions of partial differential equations of the heat-conduction type, Proc. Cambridge Philos. Soc. 43 (1947) 50-67.

[31] K.L. Kreider, Department of Mathematics, The University of Akron, Fall 2017, Advanced Numerical PDEs, http://www.math.uakron.edu/ ~kreider/anpde/penta.f

[32] A. Karma and W.J. Rappel, Phase-field method for computationally efficient modeling of solidification with arbitrary interface kinetics, Phys. Rev. E 53 (4) (1996) R3017-R3020.

[33] H.S. Udaykumar and L. Mao, Sharp-interface simulation of dendritic solidification of solutions, Int. J. Heat Mass Trans. 45 (2002) 4793-4808.

[34] Y.T. Kim, N. Goldenfeld and J. Dantzig, Computation of dendritic microstructures using a level set method, Phys. Rev. E 62 (2) (2000) 2471-2474.

[35] D. Juric and G. Tryggvason, A Front-Tracking Method for Dendritic Solidification, J. Comput. Phys. 123 (1996) 127-148.

[36] H. S. Udaykumar, R. Mittal and P. Rampunggoon, Interface tracking finite volume method for complex solid–fluid interactions on fixed meshes, Commun. Numer. Meth. Engng. 18 (2002) 89-97.

[37] R.J. Leveque and Z. Li, The immersed interface method for elliptic equations with discontinuous coefficients and singular sources, SIAM J. Numer. Anal. 31 (4) (1994) 1019-1044.

[38] G. Tryggvason, B. Bunner, A. Esmaeeli, D. Juric, N. Al-Rawahi, W. Tauber, J. Han, S. Nas and Y.J. Jan, A Front-Tracking Method for the Computations of Multiphase Flow, J. Comput. Phys. 169 (2001) 708-759.

[39] F. Gibou, R.P. Fedkiw, L.T. Cheng and M. Kang, A second-order-accurate symmetric discretization of the Poisson equation on irregular domains, J. Comput. Phys. 176 (2002) 205-227.

[40] J.A. Sethian and P. Smereka, Level Set Methods for Fluid Interfaces, Annu. Rev. Fluid Mech. 35 (2003) 341-372.

[41] F. Gibou, R. Fedkiw, R. Caflisch and S. Osher, A level set approach for the numerical simulation of dendritic growth, J. Sci. Comput. 19 (2003) 183-199.

[42] V. Slavov and S. Dimova, Phase-Field Versus Level Set Method for 2D Dendritic Growth, in: T. Boyanov, S. Dimova, K. Georgiev, G. Nikolov (Eds.), NMA 2006, Lecture Notes in Computational Science 4310, Springer-Verlag, Berlin, 2007, pp. 717-725.

[43] E. Javierre, C. Vuik, F.J. Vermolen and S. van der Zwaag, A comparison of numerical models for one-dimensional Stefan problems, J. Comp. Appl. Math. 192 (2006) 445-459.

[44] C.Min and F. Gibou, A second order accurate level set method on non-graded adaptive cartesian grids, J. Comput. Phys. 225 (2007) 300-321.

[45] M. Theillard, F. Gibou and T. Pollock, A Sharp Computational Method for the Simulation of the Solidification of Binary Alloys, J. Sci. Comput. 63 (2015) 330-354.

[46] E. Olsson and G. Kreiss, A conservative level set method for two phase flow, J. Comput. Phys. 210 (2005) 225-246.

[47] H. Chen, C. Min and F. Gibou, A numerical scheme for the Stefan problem on adaptive Cartesian grids with supralinear convergence rate, J. Comput. Phys. 228 (2009) 5803-5818.

[48] Y. Yang and H.S. Udaykumar, Sharp interface Cartesian grid method III: Solidification of pure materials and binary solutions, J. Comput. Phys. 210 (2005) 55-74.